\documentclass[%
 reprint,
superscriptaddress,
showpacs,
 amsmath,amssymb,
 aps,
prl,
floatfix,
]{revtex4-1}

\usepackage{graphicx}
\usepackage{dcolumn}
\usepackage{bm}

\usepackage{ulem}
\usepackage{xcolor}

\newcommand{\pr}{Pr$^{3+}$}

\newcommand{\thf}{$^3$H$_{4}$}

\newcommand{\odt}{$^1$D$_{2}$}

\newcommand{\LAWO}{La$_2$(WO$_4$)$_3$}

\newcommand{\PG}{\textcolor{black}} 
\newcommand{\AF}{\textcolor{black}}

\newcommand{\ML}{\color{black}}

\begin{document}


\title{Faithful Solid State Optical Memory with Dynamically Decoupled\\ Spin Wave Storage}%

\author{Marko Lovri\'c}
\altaffiliation{Present address: Sirah Lasertechnik GmbH, Heinrich-Hertz-Stra\ss e 11, D-41516 Grevenbroich, Germany}
\affiliation{%
 Technische Universit\"at Dortmund, Fachbereich Physik, D-44221 Dortmund, Germany
}
\affiliation{%
Chimie ParisTech, Laboratoire de Chimie de la Mati\`ere Condens\'ee de Paris, CNRS-UMR 
7574,\\ UPMC Univ Paris 06, 11 rue Pierre et Marie Curie 75005 Paris, France
 }%
\author{Dieter Suter}%
\affiliation{%
 Technische Universit\"at Dortmund, Fachbereich Physik, D-44221 Dortmund, Germany
}%

\author{Alban Ferrier}
\affiliation{%
Chimie ParisTech, Laboratoire de Chimie de la Mati\`ere Condens\'ee de Paris, CNRS-UMR 
7574,\\ UPMC Univ Paris 06, 11 rue Pierre et Marie Curie 75005 Paris, France
 }%
\author{Philippe Goldner}
\email{philippe-goldner@chimie-paristech.fr}
\affiliation{%
Chimie ParisTech, Laboratoire de Chimie de la Mati\`ere Condens\'ee de Paris, CNRS-UMR 
7574,\\ UPMC Univ Paris 06, 11 rue Pierre et Marie Curie 75005 Paris, France
 }%

\date{\today}

\begin{abstract}
 {\AF{We report an optical memory in a rare earth doped crystal with long storage times, up to 20 ms, together with an optical bandwidth of 1.5 MHz. This is obtained by transferring optical coherences to}} {\ML{nuclear spin coherences, which were then protected against environmental noise by dynamical decoupling.}} With this approach, we achieved a 33 fold increase in spin wave storage time over the intrinsic spin coherence lifetime. Comparison between different decoupling sequences indicates that sequences insensitive to initial spin cohrence increase retrieval efficiency. Finally, an interference experiment shows that relative phases of input pulses are preserved through the whole storage process with a visibility $\approx$1, demonstrating the usefulness of dynamical decoupling for extending the storage time of quantum memories.

\end{abstract}

\pacs{03.67.Pp, 42.50.Md, 03.67.Hk}
\maketitle

Quantum memories for light (QML) are devices capable of faithfully storing photonic quantum 
states into atomic states \cite{Lvovsky:2009fr}. Their applications include long distance quantum 
cryptography and more generally quantum networks \cite{Kimble:2008if}.  Besides atomic vapors, rare earth ions 
have recently been considered as promising candidates for solid state QMLs. This is because of 
the coherence lifetimes of their optical and nuclear spin transitions, which can reach the ms 
range \cite{Macfarlane:2002ug,Alexander:2007fz}. Moreover, these systems are well suited for 
memories with large time-bandwidth products since their optical inhomogeneous linewidth can 
exceed by several orders of magnitude the homogeneous one \PG{\cite{Macfarlane:2002ug}}. To take advantage of this 
property, the optical input signal is absorbed in an inhomogeneously broadened transition. 
\PG{Excited atomic coherences then dephase and, after a time $t$, are rephased by an optical control  pulse, resulting in an output signal at time $2t$ similar to a photon echo
\cite{Tittel:2009bp}.} Protocols like 
CRIB (Controlled Reversible Inhomogeneous Broadening) \cite{Nilsson:2005ea}, GEM (Gradient 
Echo Memory) \cite{Hetet:2008hc}, AFC (Atomic Frequency Comb) 
\cite{deRiedmatten:2008ck,Afzelius:2009gc}
or ROSE (Revival of Silenced Echo) \cite{Damon:2011tx} have been developed from this basic 
scheme to allow for high efficiency, high bandwidth and single photon level input signals. To 
reach long storage times, the optical coherence can be transferred to a ground state nuclear 
spin coherence. This is also required in the AFC protocol to obtain an on-demand memory 
\cite{Afzelius:2010fh}. Using these protocols in different rare earth crystals, recent 
demonstrations  include 1 GHz bandwidth storage \cite{Bonarota:2011bv}, 70 \% storage 
efficiency \cite{Hedges:2010dq}, entanglement storage \cite{Clausen:2011uw,Saglamyurek:2011js} and entanglement of two crystals \cite{Usmani:2012td}. 
\PG{However, the few experiments on optical to spin storage  reported storage times of only $\approx 20-50$ $\mu$s \cite{Afzelius:2010fh,Mazzera:2013vb,Timoney:2013tc}.}
In these studies, however, the spin coherence was not refocused 
and the storage time was therefore limited by the spin inhomogeneous broadening. On the other 
hand, it has been shown that very efficient control of rare earth spin decoherence can be 
achieved by combining external magnetic fields \cite{Fraval:2004cu,Lovric:2011ct} and dynamical 
decoupling with radio-frequency (RF) pulses \cite{Fraval:2005gk,PascualWinter:2012ew}.
{\ML{In demonstrations using electromagnetically induced transparency (EIT), this lead to memories with storage times up to several seconds \cite{Longdell:2005ik,Heinze:2010ej}. However, EIT is seriously limited in bandwidth because of the low oscillator strength of rare earth transitions, and the largest bandwidths reported in these memories do not exceed a few tens of kHz.
}}
Here, we demonstrate a photon echo based  memory in a rare earth 
doped crystal with a bandwidth of 1.5 MHz and storage times between 7 and 20 ms, increasing 
previous values up to three orders of magnitude. This is achieved by controlling spin 
coherence by RF dynamical decoupling (DD) sequences. Comparison between these sequences reveals that sequences insensitive to initial spin coherence increase retrieval efficiency. We finally investigated storage fidelity, a crucial point in quantum memories, and show that relative optical phases are preserved through the whole storage process. This result provides the first demonstration of high fidelity storage in an ensemble-based optical memory using dynamical decoupling.  This shows the high potential of this technique for ensemble-based memories in general.

Experiments were performed on a 0.2 at.\%  \pr:\LAWO{} single crystal. This material was 
developed for quantum memories to reach low optical inhomogeneous broadening at high \pr{} 
concentration in order to increase optical depth \cite{Beaudoux:2012ch,Goldner:2009bw}. The 
crystal  was cooled down to a temperature of $\approx$ 5 K in a cold finger liquid helium 
cryostat. Optical excitation was provided by a Coherent 899-21 dye laser stabilized to a linewidth 
$<$  20 kHz. Light propagated parallel to the crystal $b$ axis and was focused to a spot of 40 $
\mu$m inside the sample which was 4 mm thick. Optical pulse amplitude and frequency  were 
controlled by acoustic-optic modulators (AOM) in double pass {\ML{configuration, }} driven by an arbitrary 
waveform generator. The signal was detected by an avalanche photodiode. To apply radio frequency 
(RF) pulses, a 6 mm, 10 turn coil surrounded the crystal. {\ML{To reduce reflections and increase the field strength, the coil was part of a tuned circuit, which was driven with maximum RF power of 9 W and controlled by a 300 MS/s direct digital synthesizer.}}

\begin{figure}[htbp]
\begin{center}
\includegraphics[width=0.8\columnwidth]{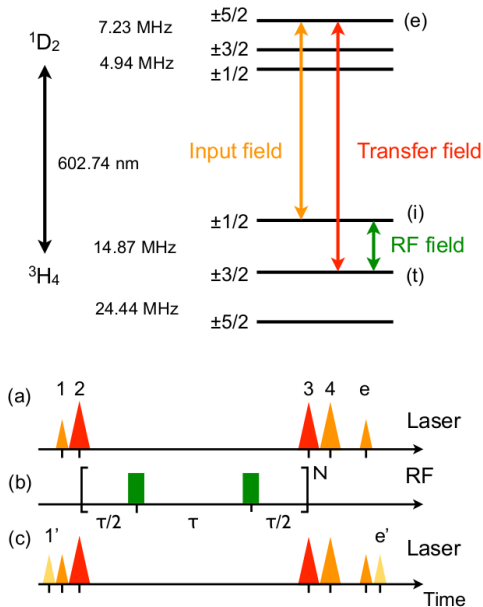}
\caption{(Color online) Upper part: hyperfine structure of \pr{} \thf{} and \odt{} levels in \pr:\LAWO{} 
and transitions used in this work. Lower part:  laser sequences for one (a) or two (c) 
pulse storage. Pulses 1 and 4 were gaussian with full width at half maximum (FWHM) lengths of 
200 and 425 ns respectively. Pulses 2 and 3 were secant hyperbolic with FWHM length of 2.25 $
\mu$s and a 2 MHz chirp. The delay $t_{12}$, between pulse 1 and 2, and $t_{34}$ were set to 2 $\mu$s.  Comparing echo 
intensity with and without transfer pulses, we deduced  a transfer 
efficiency for  fields of  87 \%   per pulse.
(b) RF pulse sequence for hyperfine transition dynamical decoupling; the basic 
block showed in bracket is repeated $N$ times (see text). }
\label{fig:scheme}
\end{center}
\end{figure}

Optical excitations were resonant with transitions between the lowest electronic level of {\ML{}the} \pr{} \thf{} 
ground and the \odt{} excited {\ML{}state multiplet} (Fig. \ref{fig:scheme}, upper part). The optical transition 
has an inhomogeneous linewidth of 10 GHz and a homogeneous one of 27 kHz ($T_2$ = 11.5 $
\mu$s). $^{141}$\pr{} has a $I=5/2$ nuclear spin and 100\% abundance. {\ML{Each}} electronic level 
has a hyperfine structure of three doubly degenerate levels (Fig. \ref{fig:scheme}, upper part) at 
zero external magnetic field. The strongest optical transition occurs between levels (i) and (e) 
\cite{GuillotNoel:2009ba} and was chosen to absorb the input signal. Spin storage  was 
performed on the (i)-(t) transition at 14.87 MHz. 

\begin{figure}[htbp]
\begin{center}
\includegraphics[width=\columnwidth]{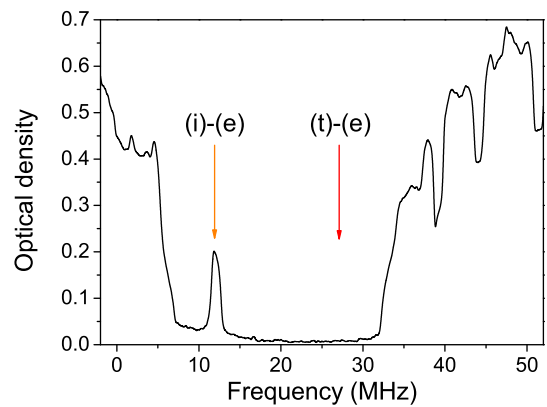}
\caption{Absorption spectrum after optical pumping.}
\label{fig:spectrum}
\end{center}
\end{figure}

As the optical inhomogeneous linewidth is much larger than the hyperfine level separation, optical pumping was first used to isolate the transitions of interest.
 {\ML{The first step of the optical pumping sequence  \cite{GuillotNoel:2009ba} consisted}} in burning a spectral pit of 25 MHz, to empty levels (i) and (t) for one class of ions. 
Population was then brought back into the (i) level of these ions by a pulse adjusted to create a 1.5 
MHz wide absorption peak. Finally, additional pulses were applied to remove unwanted spectral 
features in the pit. Fig. \ref{fig:spectrum} shows the final transmission spectrum at the end of the preparation sequence. It consists of a 
well isolated peak at 12.2 MHz corresponding to the (i)-(e) transition and a low background 
absorption on the (t)-(e) transition at 27.0 MHz. Level (t) is  therefore empty, which is required for 
efficient transfer of the optical (i)-(e) coherence to the hyperfine (i)-(t) transition. 

Fig. \ref{fig:scheme} (a) and (b) show the general memory sequence that we used. 
The input signal (pulse 1) was first 
{\ML{absorbed by}} the inhomogeneously broadened (i)-(e) transition. The resulting 
coherence was then transferred to the hyperfine (i)-(t) coherence by a $\pi$ pulse 
{\ML{resonant with}} 
the (t)-(e) transition (pulse 2). 
{\ML{The inhomogeneous dephasing of the resulting }} 
initial spin coherence was refocused by an 
even number of RF $\pi$ pulses applied to the (i)-(t) transition (see Fig. \ref{fig:scheme} (b)). 
If the pulses {\ML{are applied at a}} rate 
larger than the correlation time of the dephasing bath, the spin transition 
coherence lifetime may increase through dynamical decoupling \PG{\cite{Viola:1998ky}}. This is because  the
bath appears static, like an additional inhomogeneous broadening, between successive $\pi$ pulses. 
The RF pulses refocus this broadening and effectively increase the transition coherence lifetime. 
{\ML{Ideally}}, after 
the last RF pulse and a delay of $\tau/2$, where $\tau$ is the RF pulse separation, the hyperfine 
coherence state is the same as the initial one. Applying a second transfer at this time (pulse 3) 
brings the hyperfine coherence back to the optical domain. A final $\pi$ pulse along the (i)-(e) 
transition (pulse 4), refocuses the optical {\ML{dephasing of the}} coherence. Finally, the output signal (pulse e) appears 
as a photon echo at a time $t_{4e}=t_{12}+t_{34}$ after pulse 4, where $t_{ij}$ is the delay 
between pulses $i$ and $j$. The memory storage time is $T=t_{1e}$.


RF pulses {\ML{of rectangular shape}} were applied between the transfer pulses as shown in Fig. 
\ref{fig:scheme} (a) and (b). {\ML{Using an appropriate RF power the pulse area was set to $\pi$, while}} their duration was set to 5 $\mu$s so that their spectral width  ($
\approx$ 200 kHz) was larger than the inhomogeneous linewidth of the (i)-(t) transition (45 kHz). 
The pulse amplitude was determined by nutation and spin echo experiments. The relative phase 
of successive  RF pulses could be adjusted,  $X$ and $Y$ representing respectively 0 and 90$^
\circ$ phases in the following. {\ML{The}} optical phase {\ML{of the laser}} was independent of the RF phase, {\ML{thus for each repetition of the experiment}} the initial phase of the spins 
was 
 arbitrary compared to the RF pulse phases. Moreover, since the first transfer pulse was applied after  evolution in the optical 
domain, there was a distribution of initial spin phases. As a result, no spin echo was observed 
between RF pulses or at the end of the RF sequence. Only the final optical echo could be used 
to probe the spin coherence decay.  {\ML{The}} RF sequences consisted {\ML{of}} a basic block of length $2\tau$, {\ML{which was}} repeated  $N$ times (see Fig. \ref{fig:scheme} (b)).

We compared two dynamical decoupling sequences.
{\ML{Their building blocks, where  e.g. $\pi(\phi)$ represents a RF $\pi$ pulse with $\phi$ phase, are given by: }}
(i) $[\tau/2{-}\pi(X){-}\tau{-}\pi(X){-}\tau/2]$, Carr-Purcell-Meiboom-Gill (CPMG) sequence \cite{Meiboom:1958dq}; \newline
(ii) $[${\ML{$\mathrm{KDD}(X){-}\mathrm{KDD}(Y)$}}${-}\mathrm{KDD}(X){-}\mathrm{KDD}(Y)]$, where KDD($\phi$) =  $[\tau/2{-}\pi(\phi{+}\pi/6){-}\tau{-}\pi(\phi){-}\tau{-}\pi(\phi{+}\pi/2)-\tau${\ML{${-}\pi(\phi)$}}${-}\pi(\phi{+}\pi/6){-}\tau/2]$, Knill DD (KDD) sequence \cite{Souza:2011kd}. The CPMG sequence {\ML{has one of the highest decoupling efficiencies,}} but only when the spins are initially aligned with the rotation axis of the pulses \cite{Souza:2012ut}. The KDD sequence is designed to be insensitive to initial phases {\ML{\cite{Souza:2011kd}.}} 

\begin{figure}[htbp]
\begin{center}
\includegraphics[width=\columnwidth]{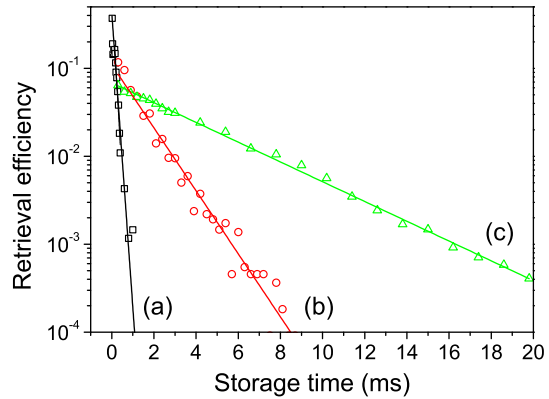}
\caption{Retrieval efficiency as a function of storage time using two RF pulses (a), KDD (b)  
and CPMG (c) dynamical decoupling sequences. The retrieval efficiency is normalized to the 
 echo intensity extrapolated at zero delay, using  
transfer pulses but no RF pulses.}
\label{fig:storage}
\end{center}
\end{figure}

Figure \ref{fig:storage} shows the output signal intensity $I$ as a function of the storage time $T$ 
for the {\ML{CPMG and KDD}} sequences.  $I$ is {\ML{normalized to}} 
the output pulse obtained with zero delay 
between transfer pulses and no RF pulses, and is plotted as a retrieval efficiency.  The pulse 
separation $\tau$ was 30 $\mu$s for KDD and CPMG sequences and was optimized for the longest 
storage time  with KDD. As shown in Fig. \ref{fig:storage}, output signal decays were 
approximately  exponential  with corresponding effective coherence lifetimes $T_{2, 
\mathrm{eff}}$ of 8.4 ms and 1.9 ms for CPMG and KDD sequences respectively. $T_{2, 
\mathrm{eff}}$ is defined by the decay of the optical echo intensity $I$ as a function of storage 
time $T$, $I=I_0\exp(-2T/T_{2, \mathrm{eff}})$. Compared to using only 2 RF pulses ($T_{2, 
\mathrm{eff}}$ = 230 $\mu$s), i.e. refocusing only the static inhomogeneous spin 
broadening, the two DD sequences significantly increase {\ML{the storage time}}  $T_{2, \mathrm{eff}}$ (see Fig. 
\ref{fig:storage}). A signal to noise ratio of 2 \PG{after 200 accumulations } is reached at a storage time $T=20$ ms for CPMG and 
at $T=7$ ms for KDD. 
CPMG provides the longest relaxation time, since it 'locks' the spins along the effective field generated by the pulses.
However, at 300 $\mu$s storage time, KDD's retrieval efficiency is nearly twice the one obtained with 
CPMG (12 \% and 6.4 \% respectively). This can be explained by the initial distribution of spin phases at 
the input of the DD sequences (see above): CPMG preserves only the component of the spin which is initially oriented along the rotation axis of the pulses,
while KDD protects all spin components and therefore the full quantum state, as required for a quantum memory \cite{Souza:2011kd}.

%


\begin{figure}
\begin{center}
\includegraphics[width=\columnwidth]{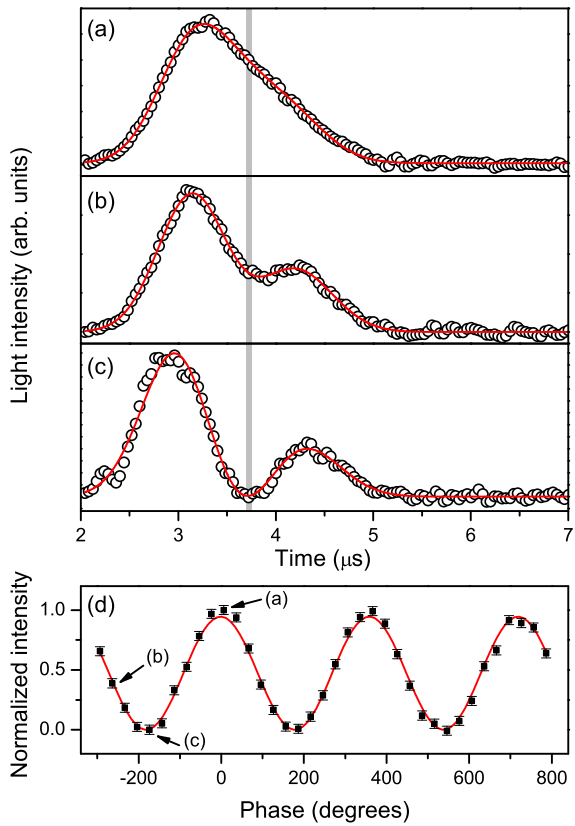}
\caption{Interfering output pulses after storage of two input pulses, with {\AF{0$^\circ$}} (a), {\AF{-270$^\circ$}} 
(b) and {\AF{-180$^\circ$}} (c) relative phases, {{\ML} for 3 ms}. Open circles: experimental data, solid line: fit using two 
gaussian pulses. {\ML{(d)}} Normalized output light intensity  averaged at 3.7 $\mu$s over 50 ns (gray area in (a)-(c)) as a
function of the relative input pulse phase. Squares: experimental data, solid line: fit with a visibility 
expression (see text). }
\label{fig:vis}
\end{center}
\end{figure}

We finally checked the fidelity of the memory by storing two optical pulses \cite{Lovric:2012tk}. 
{\ML{As}} the laser coherence lifetime is only 
about 50 $\mu$s, the phase of the output pulses {\ML{from identically repeated experiments are}} random. Only relative phases between 
successive (within $\approx 50$ $\mu$s) input and output pulses are therefore relevant.  The used sequence is shown in Fig. 
\ref{fig:scheme} (c) and (b). Compared to sequence (a), an additional gaussian pulse (pulse 1') of 200 
ns FWHM duration is stored in the memory with a delay $t_{1'1}=1$ $
\mu$s. The RF DD sequence followed the KDD scheme with $\tau = $ 30 $\mu$s  and a storage 
time of 3 ms. 
As the spectral width of the input pulses was about 5 MHz, {\ML{but the prepared absorption line ((i)-(e), see Fig.\ \ref{fig:spectrum}) was only 1.5 MHz 
wide}}, the output 
pulses e and e' were broadened to 400 ns and therefore overlapped in time and interfered. Fig. \ref{fig:vis} (a), 
(b), (c) shows the overlapping output pulses when the relative phase of the input pulses   is 0, {\AF{-270}} and 
{\AF{-180$^\circ$}} respectively. 
The output intensities 
were well modeled by two overlapping gaussian pulses (see Fig. \ref{fig:vis} (a), (b), (c)), confirming the origin of the output signal variations.
Light intensity was then averaged over 50 ns around the 
center of the interfering region (see Fig. \ref{fig:vis} (a), (b), (c)) and normalized. This intensity $I_n(\phi)$ is plotted against {\ML{the}} input pulses relative phase $\phi$ in 
Fig. \ref{fig:vis} (d) and was well fitted by {\ML{the}} 
visibility expression  $I_n(\phi) = (I_{\mathrm{max}}/2)(1+V\sin(\phi))$ with $V=0.99$. 
Relative phases are therefore preserved  through the whole storage process, which is a key 
requirement for a quantum memory. Similar results were obtained with the CPMG sequence 
with storage times up to 10 ms. 

In conclusion, we have demonstrated {\ML{the feasibility of coherence transfer to spin states and control by dynamical decoupling to drastically extend the storage 
time in a high bandwidth}} photon echo based optical memory. 
\PG{Depending on the decoupling sequence, storage times between 7 and  20 ms have been 
achieved together with an optical bandwidth of 1.5 MHz. }
We found that decoupling sequences insensitive to 
initial spin states, such as KDD, increase the retrieval efficiency. Finally, we demonstrated that the relative phase of input pulses was preserved through the whole storage process, as was shown by storing two input pulses which are allowed to interfere at the output of the memory. {\PG{This key property allows considering application of these techniques to extend storage times of ensemble-based quantum memories.}}
%

\begin{acknowledgments} The authors thank Mikael Afzelius for useful discussions.
This work was supported by the European Union FP7 projects QuRep (247743) and CIPRIS 
(Marie Curie Action - 287252).

\end{acknowledgments}

%
%


\end{document}